\documentclass[twoside]{ilcws10}
\usepackage[latin1]{inputenc}
\usepackage[dvips]{graphicx,epsfig,color}
\usepackage{wrapfig,rotating}
\usepackage{amssymb,amsmath,array}

\def\ie{{\it i.e.}}
\def\eg{{\it e.g.}}

\def\etal{{\it et al.}}

\begin{document}


\title{Light Squarks and Gluinos at TeV-scale $e^+e^-$ Colliders 
} 
\author{Thomas G. Rizzo 
\thanks{Work supported by Department of Energy contract DE-AC02-76SF00515.}
\vspace{.3cm}\\
SLAC National Accelerator Laboratory \\
2575 Sand Hill Rd., Menlo Park, CA, 94025, USA
}

\maketitle

\begin{abstract}
In the general MSSM, first and second generation squarks and gluinos may be sufficiently light to be produced and 
studied at $e^+e^-$ colliders operating in the 0.5-1 TeV energy range. After a reminder that the MSSM is {\it not} 
the same as mSUGRA, we provide a brief overview of these possibilities within this more general framework.
\end{abstract}

\section{Introduction}

\begin{figure}[htbp]
\centerline{\includegraphics[width=7.5cm,angle=0]{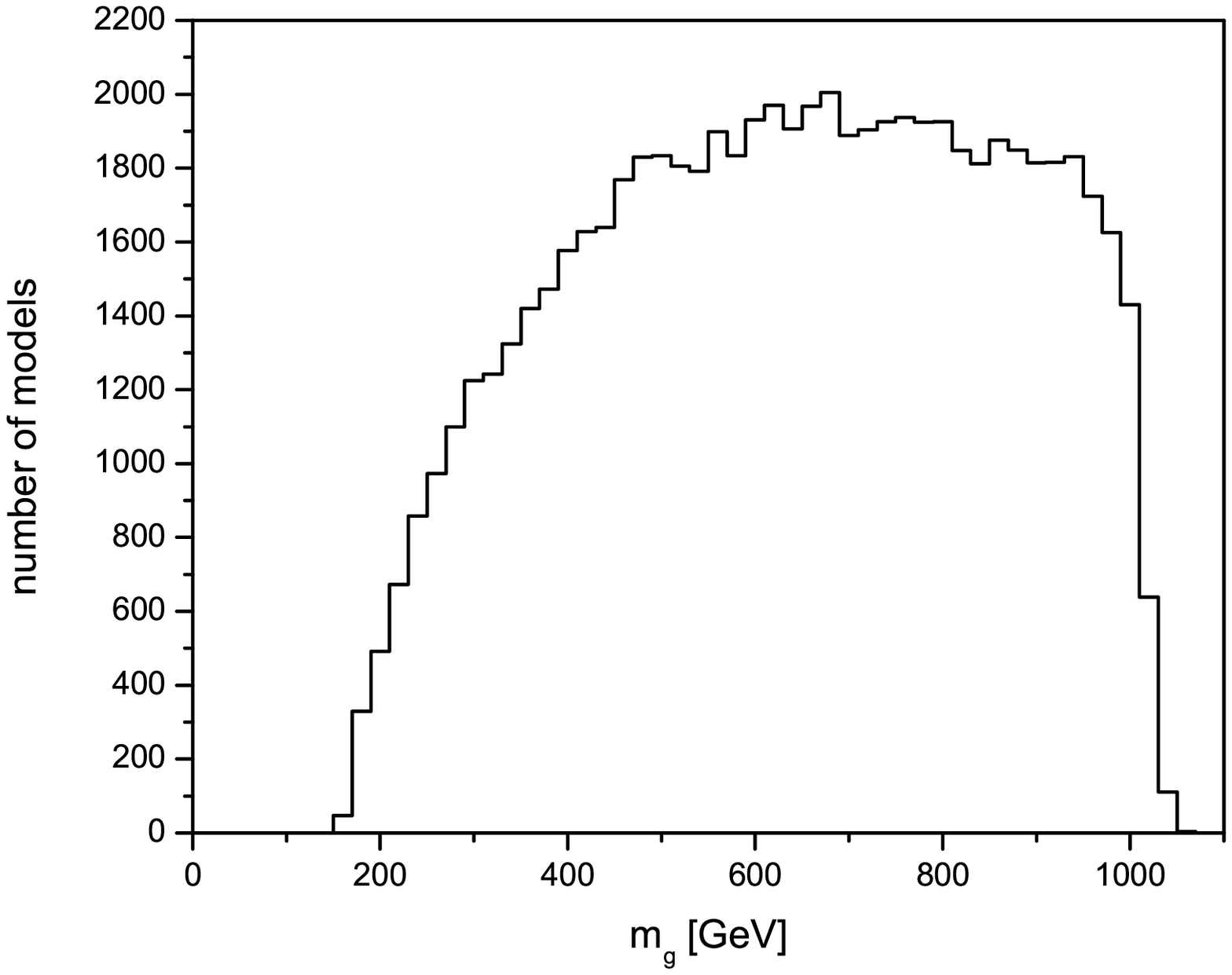}
\hspace{0.1cm}
\includegraphics[width=7.5cm,angle=0]{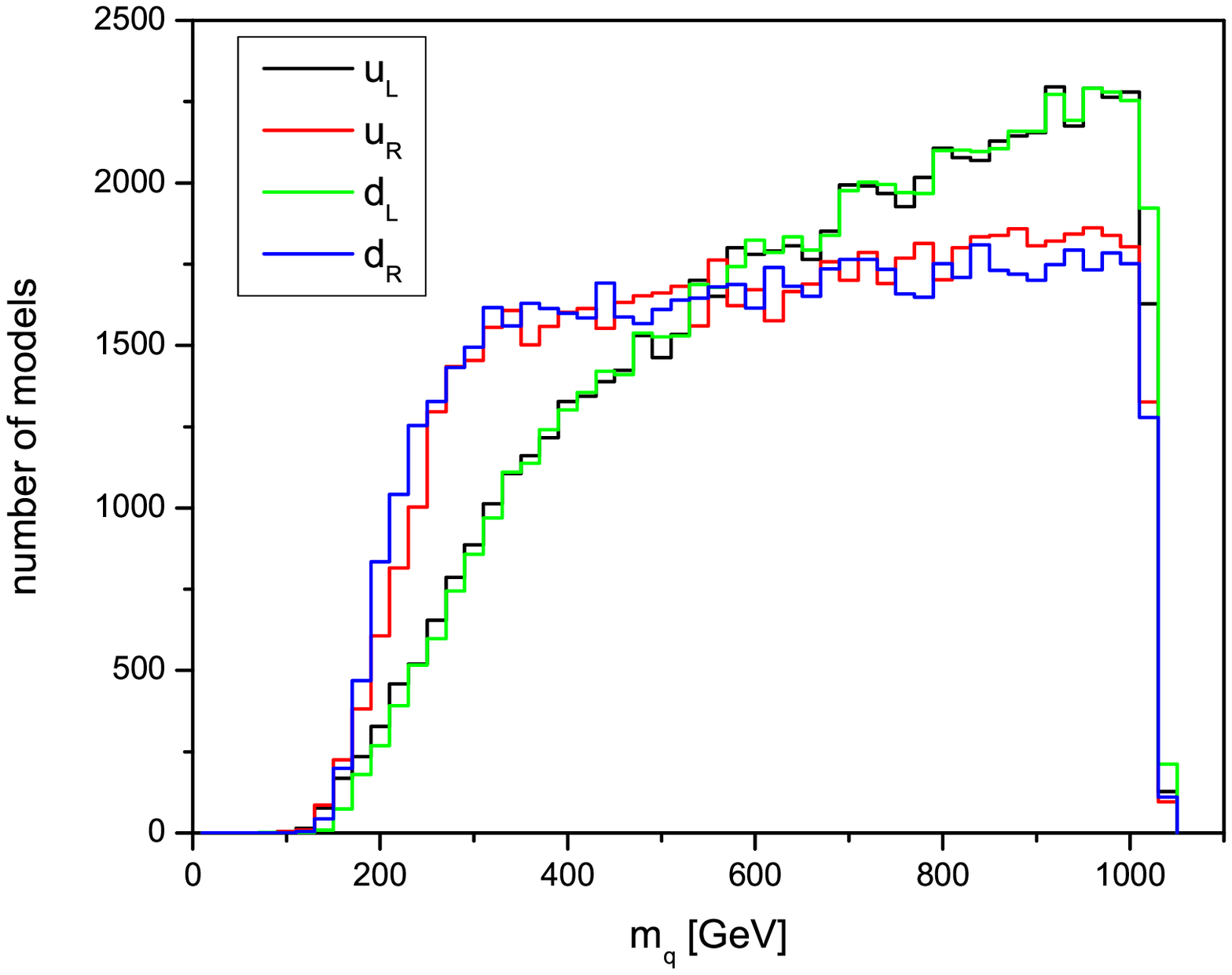}}
\vspace*{0.0cm}
\caption{Distribution of gluino and squark masses resulting from the flat prior parameter scan of the pMSSM as given by the 
analysis in Ref.{\cite{FEATURE}}.}
\label{fig0}
\end{figure}

Future linear colliders will be very useful for the detailed exploration of the new physics we expect 
to be discovered at the LHC. Supersymmetry is certainly one of the leading candidates for such new physics. 
However, even its simplest manifestation, the MSSM, is impossible to study in all its generality due to 
the $\sim 100$ free parameters associated with soft SUSY breaking. For this reason the MSSM has usually 
been studied within certain specific SUSY breaking frameworks, such as mSUGRA, since there are then only 
a few independent parameters to explore. While these SUSY breaking mechanisms are all theoretically well-motivated 
they are also necessarily highly constraining and so one can wonder how representative they are 
of the more general MSSM and whether or not the conclusions we draw from studying them can always be trusted in 
all situations. For example, it well known within mSUGRA that searches for squarks and gluinos at 
the Tevatron place rather strong lower bounds on their masses in the range of $\sim 350-400$ GeV{\cite {CDF,D0}}. These results 
are predicated on the assumption that in mSUGRA scenarios the production of squarks and gluinos and their 
subsequent decay lead to multiple high $p_T$ jets as well as significant MET due to the rather large 
mass splitting between the squarks/gluinos and the LSP (which is typically the lightest neutralino). 
These and similar results have motivated several sets of benchmark points{\cite {SPS,ATLAS,CMS}} for collider 
studies all of which having squarks and gluinos with masses in excess of $\sim 500$ GeV and thus beyond 
the range accessible to even a 1 TeV $e^+e^-$ collider. However, one can easily imagine other SUSY breaking 
scenarios wherein these mass splittings are much smaller. In such cases the resulting jets from decays are too soft to pass 
the analyses cuts causing the squarks and gluinos to be missed at the Tevatron{\cite {Jay}}. This and others scenarios leading 
to the possibility of light 1st and 2nd generation squarks and gluinos can been realized in more general models as discussed in 
Ref.{\cite {FEATURE}} with 
the results being shown in Fig.~\ref{fig0} for the case of a flat prior scan. (Similar results are obtained in the case of a log 
prior scan.) By obvious extension one can imagine constructing scenarios wherein the 
conventional inclusive MET analyses at the LHC end up missing SUSY in some cases as has been shown in Ref.{\cite {FEATURE2}}. As 
these analyses demonstrate, the possibility of light squarks and gluinos remains a viable one that should be examined in detail at 
$e^+e^-$ colliders. Here, we make a few comments on these possibilities, pointing out that much work remains to be done if such 
scenarios are actually realized at future $e^+e^-$ colliders.

\begin{figure}[htbp]
\centerline{\includegraphics[width=4.5cm,angle=90]{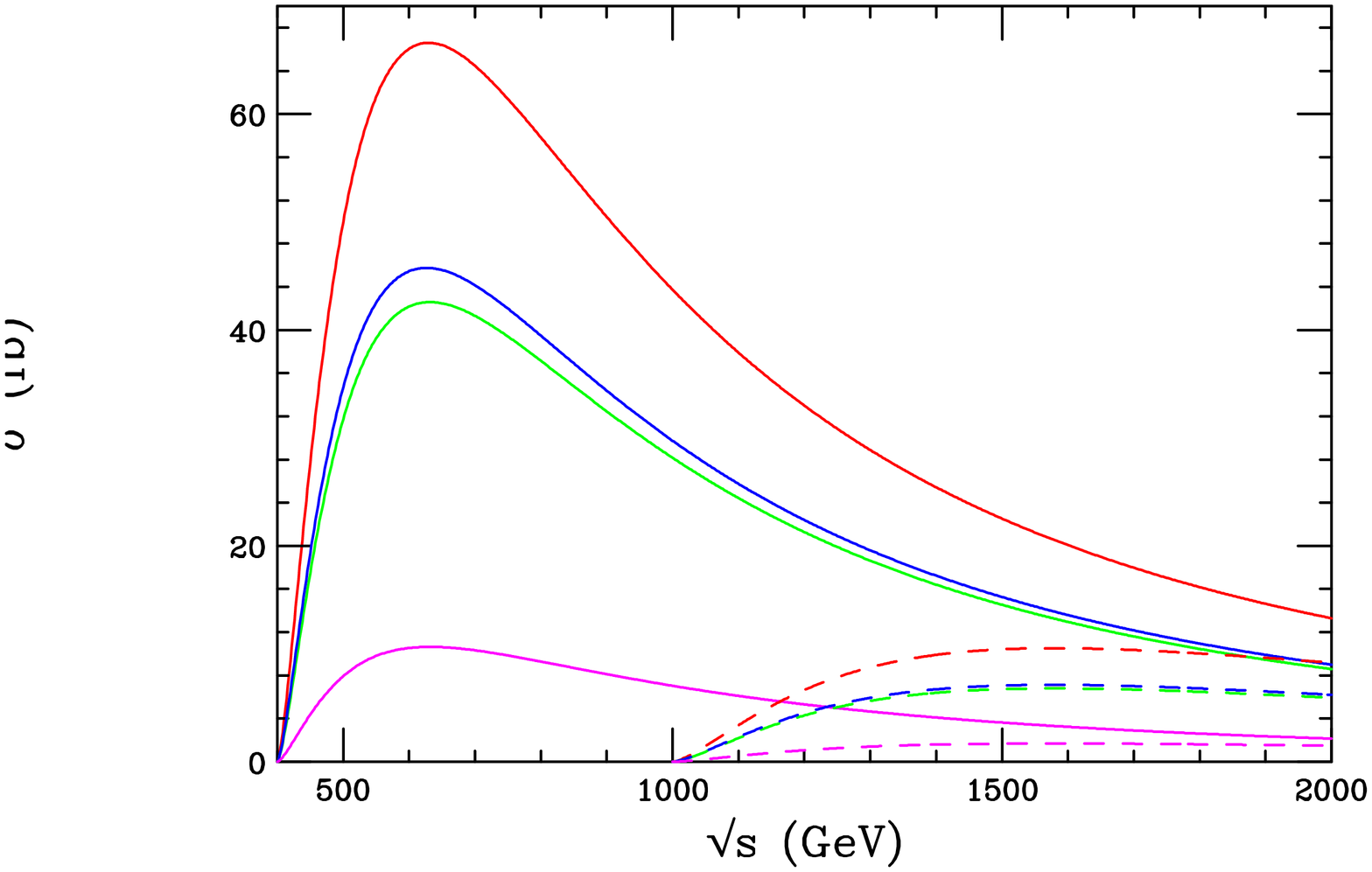}
\hspace{0.1cm}
\includegraphics[width=4.5cm,angle=90]{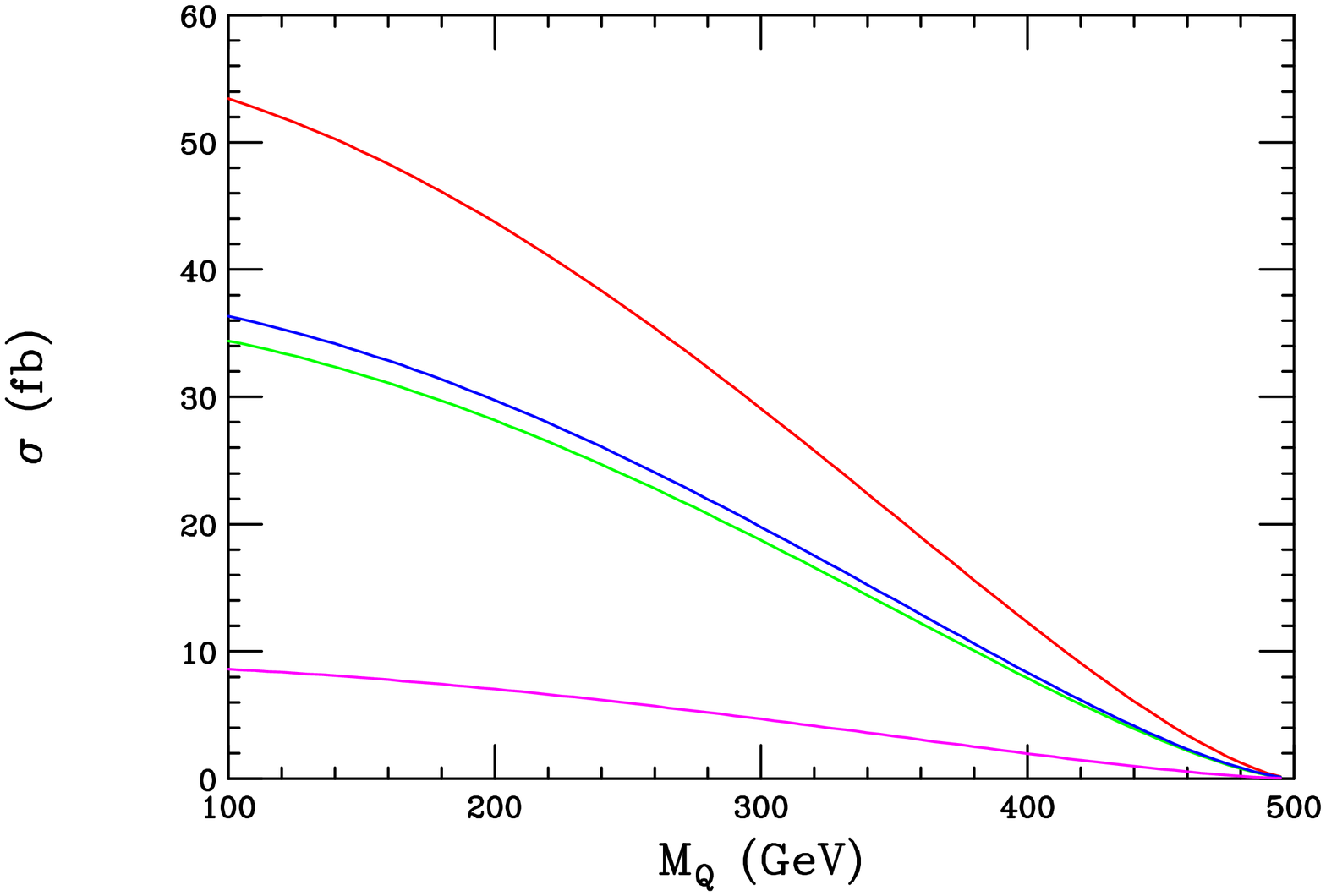}}
\vspace*{0.0cm}
\caption{Squark pair production as a function of $\sqrt s$ at $e^+e^-$ colliders for masses of 200 and 500 GeV. From top 
to bottom the curves correspond to $\tilde {u_L}, \tilde {d_L}, \tilde {u_R}$ and $\tilde {u_L}$, respectively in the LH 
panel. Same in the RH panel but now as a function of the squark mass with $\sqrt s=1$ TeV held fixed.}
\label{fig1}
\end{figure}

\section{Squarks and Gluinos in $e^+e^-$ Collisions}

The simplest signals for squark/gluino production at $e^+e^-$ colliders are jets plus MET with the number of jets possibly indicative of  
the mass ordering of the squarks and gluinos at LO in SQCD.{\footnote {As shown in Ref.{\cite {FEATURE,FEATURE2}} all such orderings are 
essentially equally likely.}} For example, if $M_{\tilde q}> M_{\tilde g}$ we may have the decays such as $\tilde q \to q\tilde g$ 
and $\tilde g \to q\bar q \chi$ whereas if $M_{\tilde q} < M_{\tilde g}$ then instead we may have the decays $\tilde g \to q\tilde q$ and 
$\tilde q \to q\chi$. Of course, the gluinos may lie in the middle of the squark mass spectrum as well since some squarks may be significantly 
heavier than others. In principle, all of the $\tilde q \tilde q^*, \tilde q \tilde g$ and $\tilde g \tilde g$ final states may be 
accessible at $e^+e^-$ colliders at some level depending on the machine energy, integrated luminosity and the details of the sparticle 
spectrum. The actual signatures for squark and gluino production critically depend upon how the gaugino state(s) $\chi$ themselves decay 
subsequent to their production in the squark/gluino decay chain. In the absence of Yukawa couplings for the first 2 generations, the states 
$\chi$ can only have bino (and/or wino) components for decays arising from RH(LH)-squarks and can in the simplest case lead to, \eg,  
ME {\it provided} $\chi$ is identified with the LSP. Of course, in general, $\chi$ need not be the LSP or even neutral  
and so the resulting decay chain can be much more complex. For example, $\chi$ could be a long-lived chargino so that SUSY events no longer 
have any ME in their final states or $\chi$ could be a heavier neutralino that lives sufficiently long to decay to a non-pointing photon inside 
the detector volume. Furthermore, $\chi$ may decay to other SUSY particles that can have unusual decay patterns themselves depending 
upon the identity of the nLSP. In fact, in the more general MSSM, the nLSP can be any of the 13 possible states including the squarks and gluinos 
themselves{\cite {FEATURE2}} which allow them to possibly be long-lived. Clearly, outside of the presence of multijet activity, the final states 
in squark and gluino decays may be quite complex and highly model dependent. 

\begin{figure}[htbp]
\centerline{\includegraphics[width=9.0cm,angle=0]{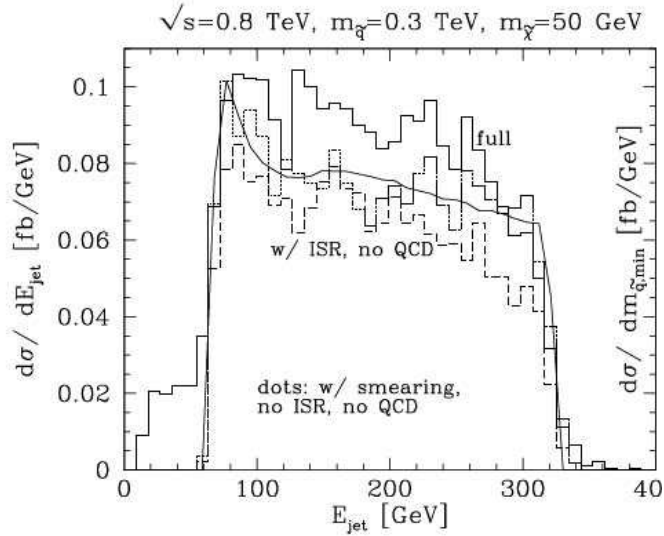}}
\vspace*{0.0cm}
\caption{Typical jet energy spectrum from squark decay to the LSP from the analysis of Drees \etal ~in Ref.{\cite {drees}}.}
\label{fig2b}
\end{figure}
\begin{figure}[htbp]
\centerline{\includegraphics[width=4.5cm,angle=90]{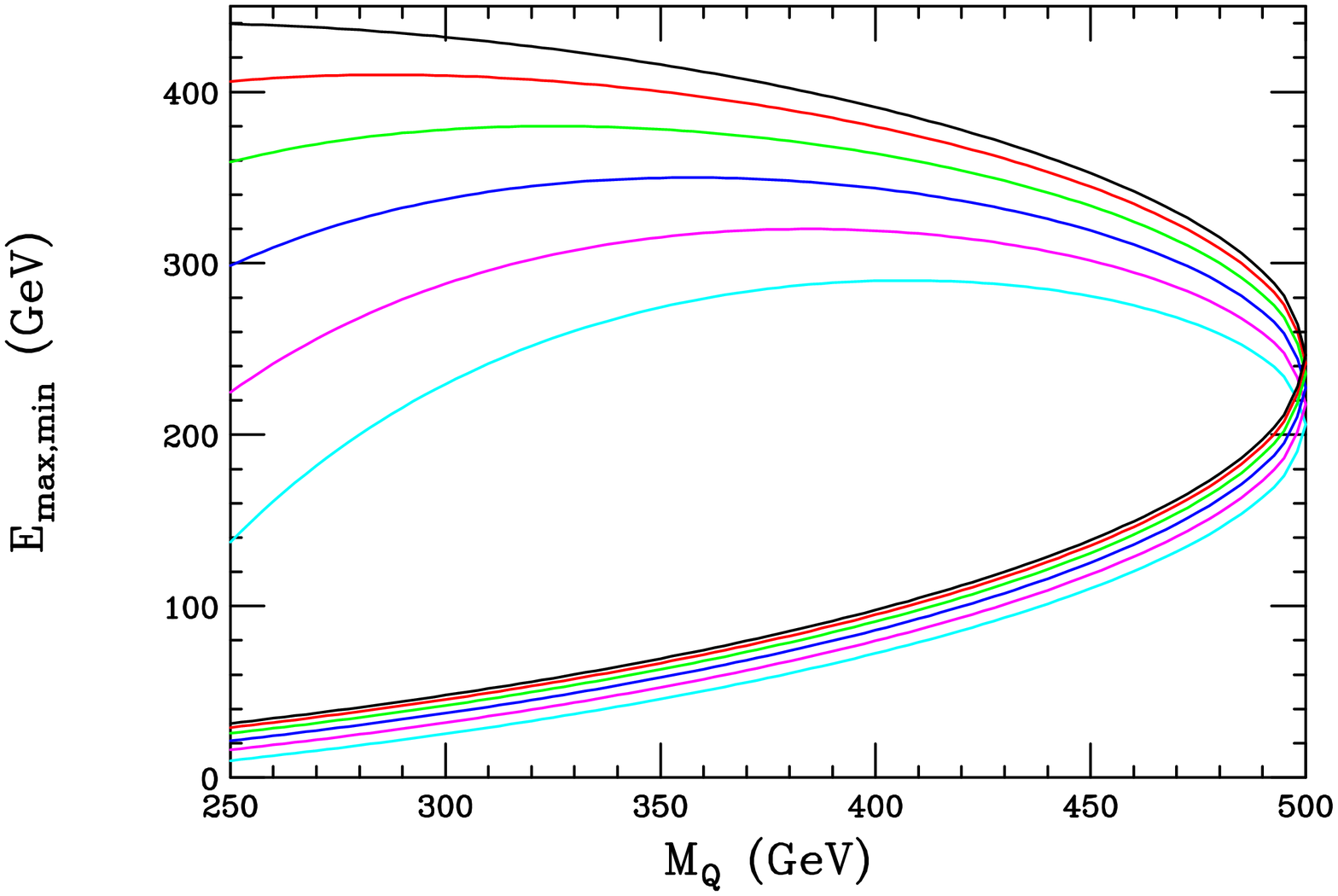}
\hspace{0.1cm}
\includegraphics[width=4.5cm,angle=90]{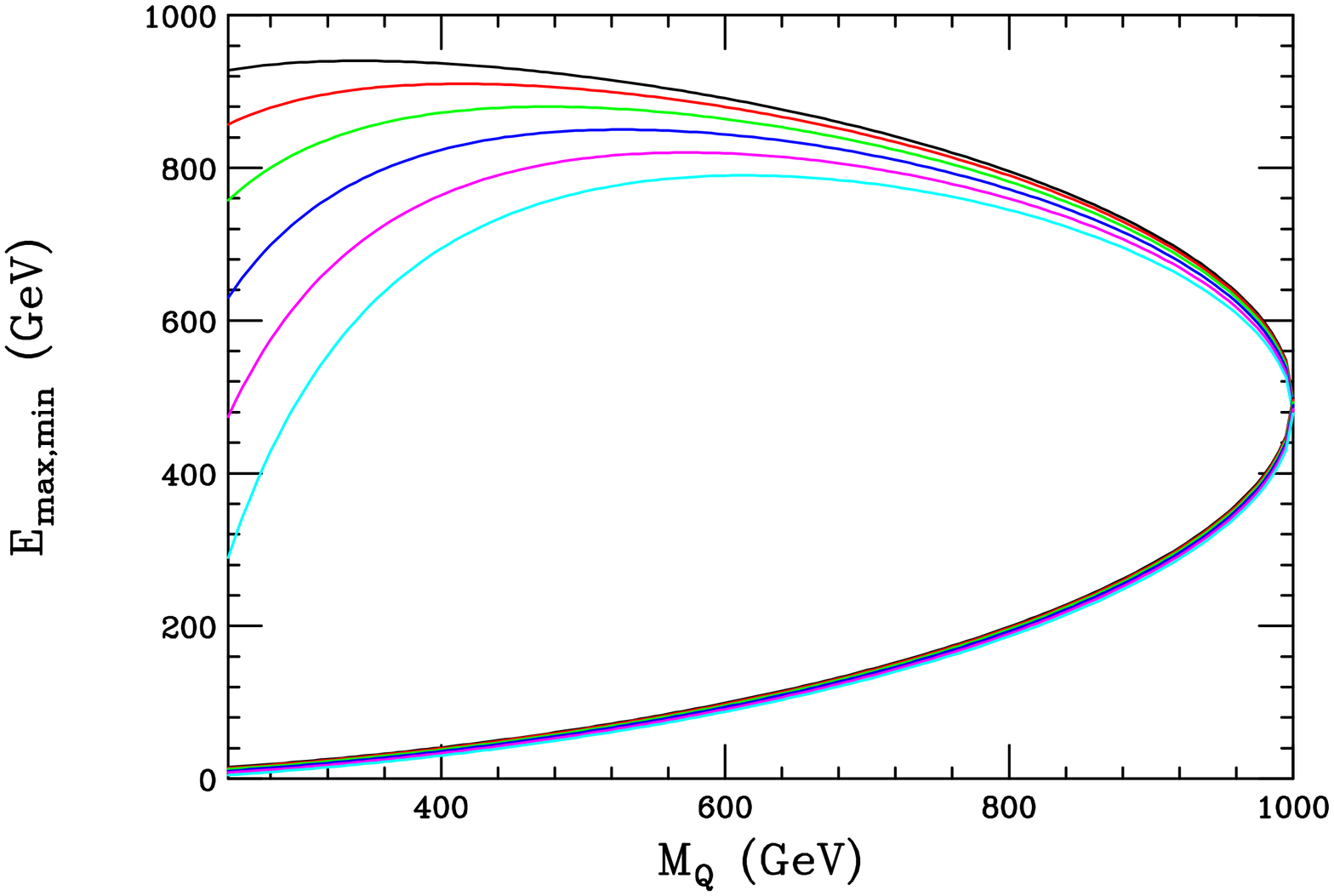}}
\vspace*{0.0cm}
\caption{Minimum and maximum jet energies as a function of the squark mass at $\sqrt s=1$(left) and 2(right TeV for various 
assumed LSP masses assuming the 2-body decay to the LSP. Here, top to bottom in the top of each panel, the curves correspond 
to LSP masses of 60 to 210 GeV in 30 GeV steps.}
\label{fig2}
\end{figure}

Squark pair production is the simplest process to study and has a standard spin-0, $s$-channel cross-section behavior, \ie, 
$d\sigma \sim \beta^3 \sin^2\theta$ with typical rates shown in Fig.~\ref{fig1}. In particular note the rather slow $\beta^3$ 
turn-on of this production process. $e^\pm$ beam polarization can used to help distinguish LH from RH squarks via their electroweak 
couplings to the $Z$. The squark masses may be determined via the usual threshold measurement techniques or by reconstructing 
specific final states. For example, as is well-known, if the squark and anti-squark each decay directly to j+LSP, then measurements 
of the endpoints and edges of the jet energy spectra{\cite {drees}} can be used to extract the squark and LSP masses as can be 
seen in Figs.~\ref{fig2b} and ~\ref{fig2}. SM backgrounds to this process from $\gamma \gamma$ initial states and from $W,Z\to jj$ 
can be mostly removed by employing jet acoplanarity, dijet mass and large ME cuts. One should be 
reminded, however, that the SUSY spectrum can be rather complex with many sparticle states being produced simultaneously. This can 
lead to many additional simultaneous sources of high energy jet spectrum edges so this simple analysis may be rather more complex than 
is usually advertised. 
Furthermore, since the squarks in the first two generations of a given 'handedness' and charge are likely to be degenerate to a 
very high degree (with splittings at most at the level of $\sim$MeV) flavor tagging will 
be useful for, \eg, separating out jets from scharm decay from those arising from $\tilde u$. Thus even this simplest final state may 
require a detailed study to completely understand. Of course if the $\chi$ is not the LSP then reconstructing the entire final state 
from squark decays will prove to be even more difficult.

\begin{figure}[htbp]
\centerline{\includegraphics[width=4.5cm,angle=90]{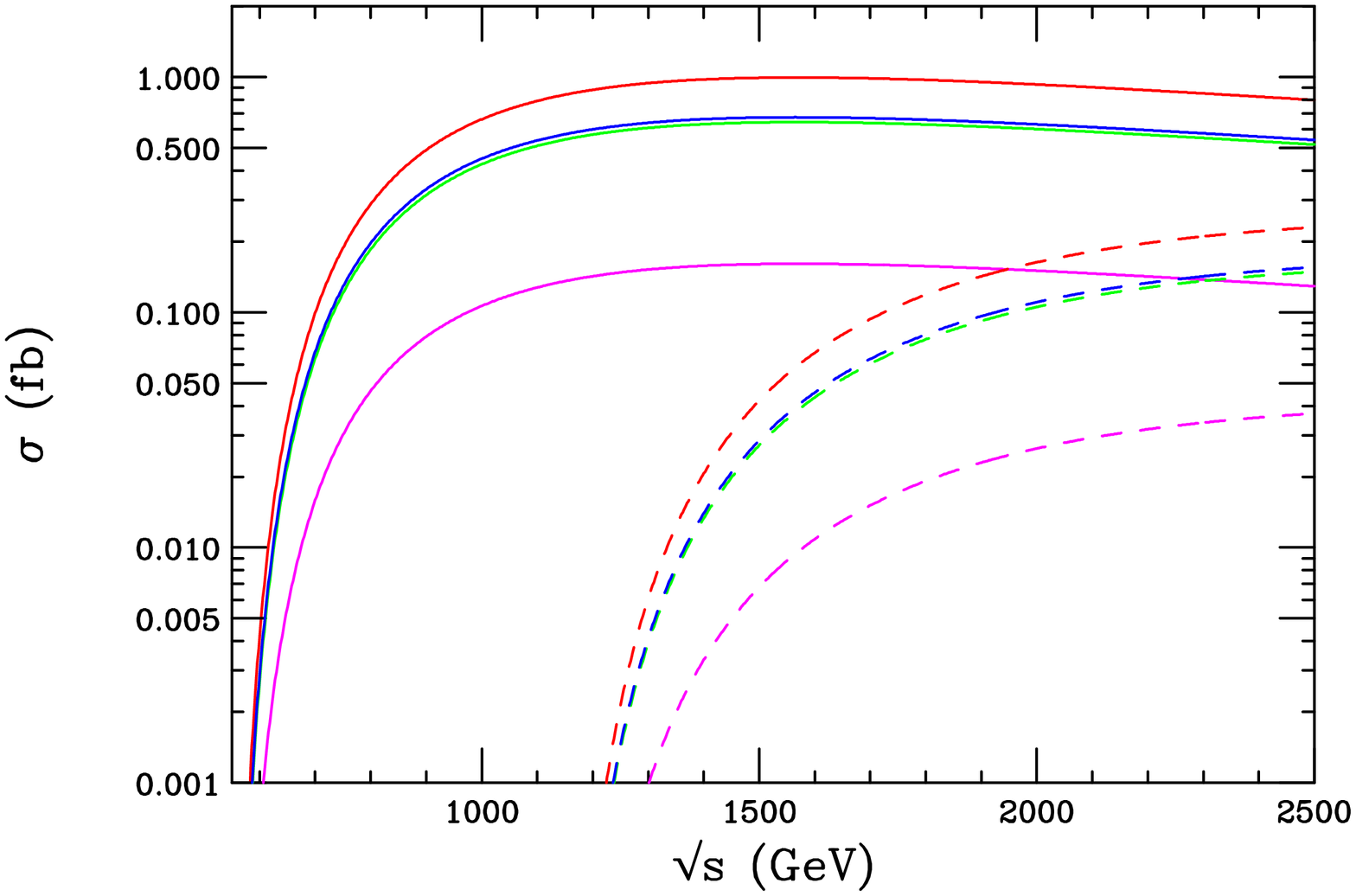}
\hspace{0.1cm}
\includegraphics[width=4.5cm,angle=90]{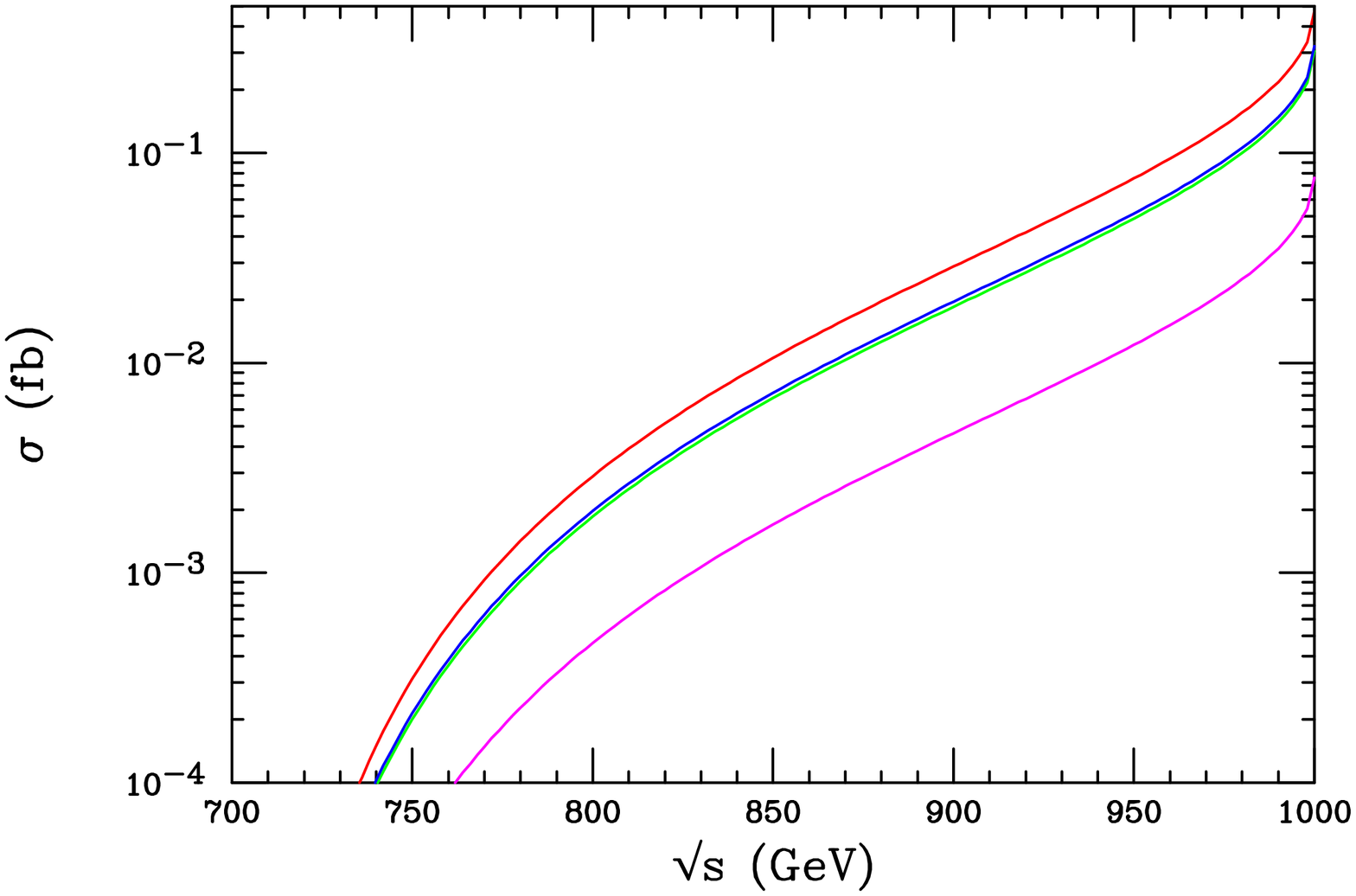}}
\vspace*{0.0cm}
\caption{Squark-gluino associated production as a function of $\sqrt s$ for SPS1a(dashed) and for the case of 250(300) 
squarks(gluinos) in the LH panel and for 500(200)squarks(gluinos) in the RH panel. The squark flavor labels are as in Fig.1.}
\label{fig3}
\end{figure}

If squarks are sufficiently long-lived due to spectrum degeneracies the threshold region may be more complex than discussed above 
due to the formation of squarkonium 
bounds states. These have been studied in $e^+e^-$ collisions to some extent in the case of stops{\cite {Fabiano01}} but not for the squarks of  
the first two generations which can have somewhat different phenomenology. . 

Perhaps the most direct way to access gluinos in $e^+e^-$ collisions, provided they are heavier than the squarks, is production in association with 
squarks, \ie, via  the process 
$e^+e^- \to \tilde g \tilde q q${\cite {Brandenburg08}}. {\footnote {Clearly if gluinos are lighter than squarks they can be 
accessed directly through squark decays.}} Typical cross sections for this process are shown in Fig.~\ref{fig3}. Note these cross section values, 
somewhat below $\sim 1$ fb,  are substantially smaller than in the typical case of squark pair production, \ie, $\sim 20-50$ fb. This is not 
completely unexpected since associated production is a 3-body process with an additional QCD coupling being 
present. Clearly, to access gluino properties 
using this mechanism will require substantial integrated luminosity. Note the rapid rise in the cross section in the RH panel which arises when 
the intermediate squark goes on-shell.   

\begin{figure}[htbp]
\centerline{\includegraphics[width=15.0cm,angle=0]{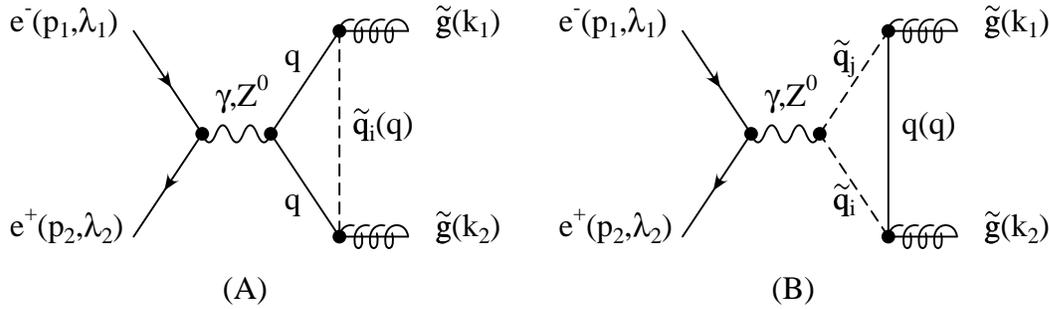}}
\vspace*{0.0cm}
\caption{Diagrams contributing to gluino pair production in $e^+e^-$ annihilation from Ref{\cite {BergeKlasen}}.}
\label{fig4}
\end{figure}

Of course, it is also possible to produce a pair of gluinos directly via their couplings to virtual photons and $Z$'s through loop diagrams 
involving both squarks and quarks as shown in Fig.~\ref{fig4}. In general,   
the cross section for gluino pair production in $e^+e^-$ collisions can be written in the form $d\sigma \sim [A(1-4\lambda_1\lambda_2)+
B(2\lambda_1-2\lambda_2)]~\beta^3~(1+\cos^2 \theta)$ whose basic structure follows directly from the fact that the gluinos are spin-1/2 
Majorana fermions{\cite {BergeKlasen}} produced by $s-$channel exchanges. Here, the $\lambda_i$ label the polarizations of the incoming 
$e^\pm$ and $A,B$ result from summing over a large number of 1-loop amplitudes. In general, the calculation of the coefficients $A,B$ 
can involve the entire strongly interacting sector of the MSSM so that the resulting cross sections can be very sensitive to the bulk 
of the MSSM parameter space. This being the case, understandably, up to now the $e^+e^- \to \tilde g \tilde g$ process has mostly been 
studied within the constraints of the mSUGRA scenario with its limited free parameter set. Studies of this process in a more general context 
would be very beneficial.
\begin{figure}[htbp]
\centerline{\includegraphics[width=6.5cm,angle=0]{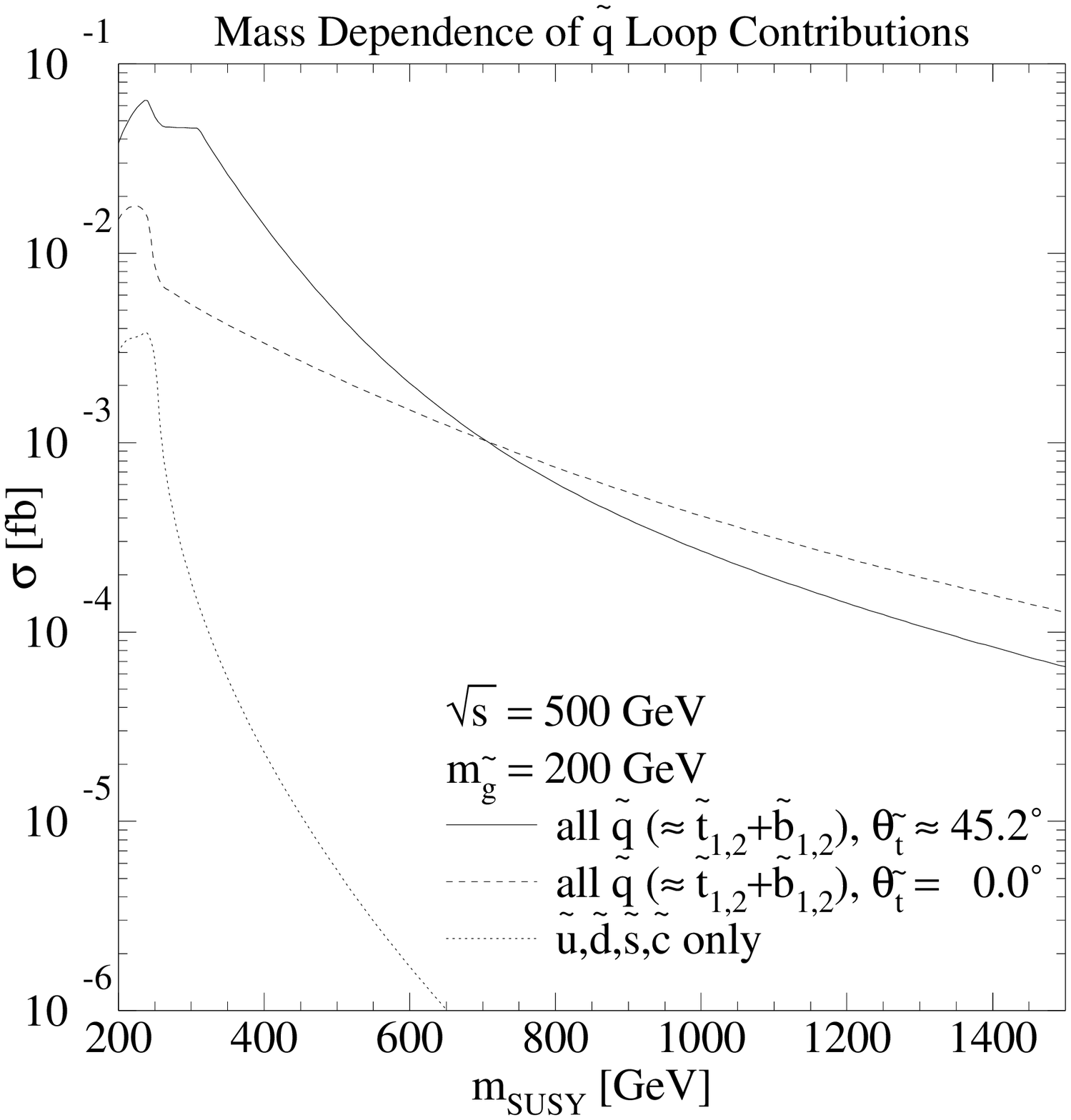}
\hspace{0.1cm}
\includegraphics[width=6.5cm,angle=0]{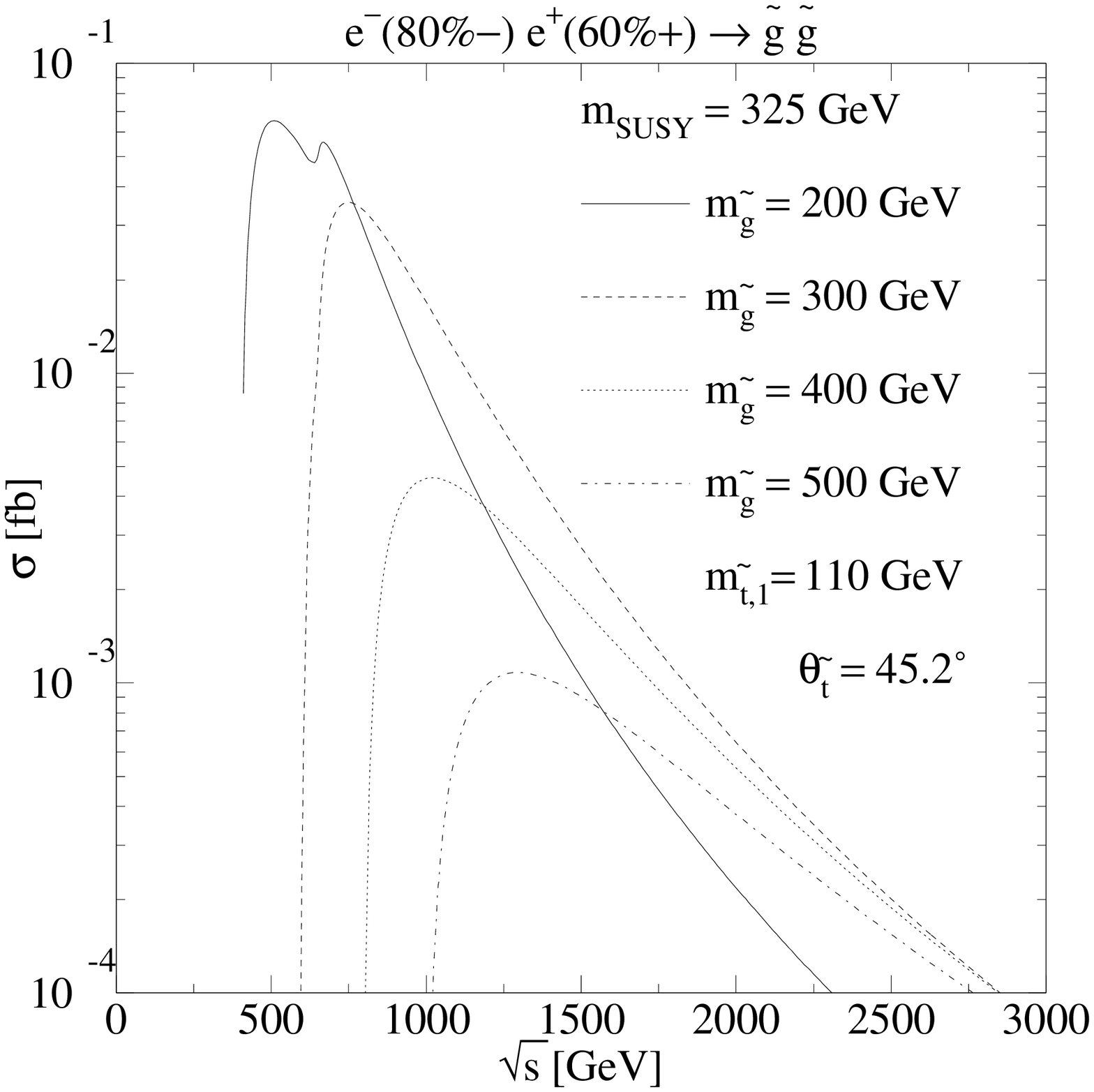}}
\vspace*{0.0cm}
\caption{Gluino pair production cross sections as a function of the SUSY scale for fixed $\sqrt s$ and as a function of $\sqrt s$ 
for different MSSM parameters as given in Ref.{\cite {BergeKlasen}}.}
\label{fig5}
\end{figure}

As discussed in detail in Ref.{\cite {BergeKlasen}}, significant rates for this process are more favorable if squark mass degeneracies, 
which are common in mSUGRA, can be removed from the spectrum. The main reason for this is that the contributions of LH- and RH-squarks 
tend to cancel as do those arising from within a given LH-squark doublet due to their opposite value of $T_3$. Furthermore, larger 
rates are favored if the squark masses take on somewhat smaller values. Within the mSUGRA scenario the dominant contribution clearly 
arises from stop loops since mixing can reduce their mass and split the LH and RH mass states. This stop dominance need not be the case 
in the more general MSSM where the various squark masses and mixings can be uncorrelated but this situation has not yet been fully examined. 

\begin{figure}[htbp]
\centerline{\includegraphics[width=6.5cm,angle=0]{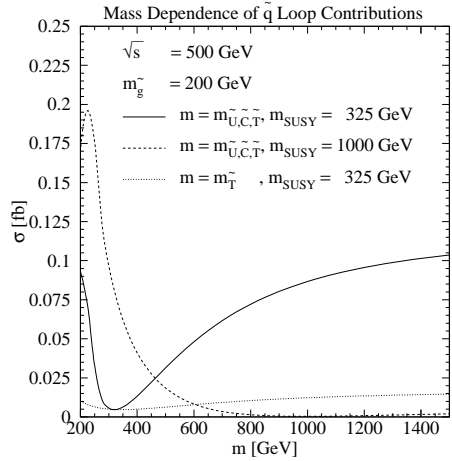}}
\vspace*{0.0cm}
\caption{Gluino pair production cross sections as a function of the squark masses for fixed $\sqrt s$ as given in 
Ref.{\cite {BergeKlasen}}.}
\label{fig6}
\end{figure}

Figs.~\ref{fig5} and ~\ref{fig6} from Ref.{\cite {BergeKlasen}}show typical gluino pair production cross sections in the case of mSUGRA 
for a range of model parameters. Note, however, that the typical values one obtains lie in the range $\sim 0.1-1$ fb, comparable to that 
obtained from the associated squark-gluino production process discussed above. However, it is not obvious that somewhat large gluino pair production 
cross sections could not be obtained for suitable chosen ranges of the general MSSM parameters. 

\begin{figure}[htbp]
\centerline{\includegraphics[width=11.0cm,angle=0]{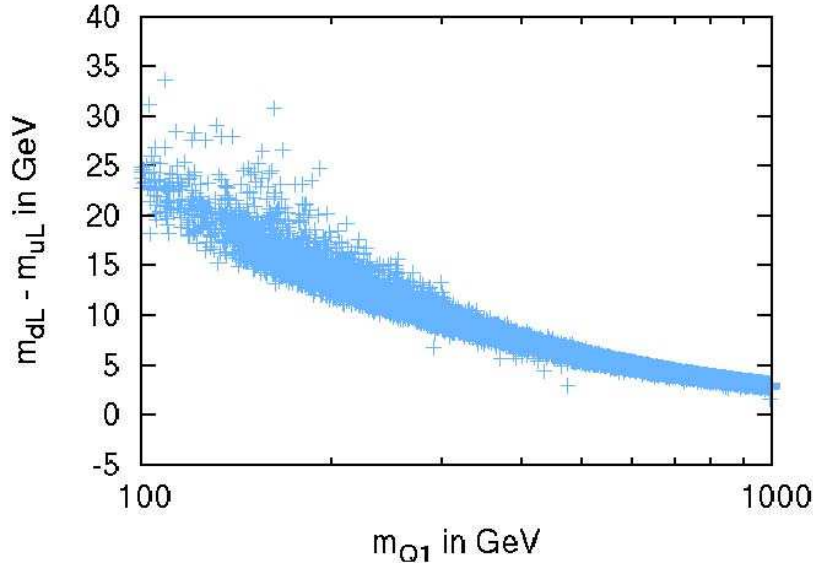}}
\vspace*{0.0cm}
\caption{Mass splitting between $\tilde {u_L}$ and $\tilde {d_L}$ including loop corrections as a function of the weak scale squark mass
parameter.}
\label{fig7}
\end{figure}

As is well-known{\cite {LHCLC}} the LHC determination of squark masses will likely not be much better than at the level of 
$\sim \pm 15-20$ GeV although relative mass differences will be much better determined. Even with the data made available from 
a $\sqrt s$=500 GeV linear collider, given the assumption mSUGRA and making use of the corresponding mass relationships, this only reduces 
these mass uncertainties to the $\sim \pm 10-15$ GeV level{\cite {LHCLC}}. The reason for this poor showing is the 
assumption that the squarks will be too massive to be pair produced at such a low value of $\sqrt s$; if squarks are in fact accessible these 
numbers might improve by up to an order of magnitude. One of the things we'd like to determine well is the relative mass splitting between the 
$\tilde {u_L}$ and $\tilde {d_L}$ as soft-SUSY breaking implies that this should be completely determined by electroweak symmetry breaking 
effects at tree-level. In fact, one finds $M_{d_L}^2-M_{u_L}^2= M_W^2 (\tan ^2 \beta-1)/(\tan ^2 \beta+1)$, with $\tan \beta$ being the ratio 
of the two Higgs doublet vevs; an identical tree-level result exists for the splitting for the LH selectron and the sneutrino. A probe of our 
understanding of SUSY would be to test this relationship as well as possible. Of course, once loop corrections are included this mass 
splitting becomes somewhat more complex, as is shown in Fig.~\ref{fig7} as a function of the TeV scale soft breaking squark mass parameter, 
but the generally expected behavior pattern is maintained.  Here we see that ($i$) over most of parameter space $\tilde {u_L}$ is less massive 
than $\tilde {d_L}${\footnote{This need not be the case if the squark soft mass is low and the loop corrections are of the same magnitude.}}, 
($ii$) this splitting decreases rapidly as the soft squark mass parameter grows and that ($iii$), for a given value of this parameter there 
is a reasonably narrow range of possible mass splitting values. From the tree-level expression we know that this predicted mass splitting 
primarily reflects the $\tan \beta$ dependence displayed above but also show many other, but weaker, MSSM parameter dependencies 
through the one-loop mass corrections. One might imagine that $e^+e^-$ measurements may be able to determine this splitting at the $\sim 1$ 
GeV level. 

\section{Summary and Conclusions}

Squarks and gluinos may be light enough to be produced at an $e^+e^-$ collider operating in the 0.5-1 TeV energy range in 
non-mSUGRA SUSY breaking scenarios. The decays of such particles, though always leading to jets, may or may not produce 
ME depending on the nature of the SUSY spectrum. The electroweak properties of the squarks can be determined using polarized 
beams and production cross section measurements. The simultaneous production of essentially degenerate 1st and 2nd generation 
squarks can be problematic although efficient charm tagging would be helpful. If (some) squarks are heavier than gluinos then 
gluinos may be best studied as squark decay products. If gluinos are heavier than squarks, associated production can lead to 
an unambiguous, although small, production cross section somewhat below $\sim 1$ fb. Gluino pair production is perhaps more 
interesting as the cross section for this process involves the entire strongly interacting sector of the MSSM. The rate is 
expected to be rather small for this process in the case of mSUGRA but has not been well-studied in the more general MSSM. 

Hopefully the LHC will soon tell us if SUSY is realized in nature.

\section{Acknowledgments}
Work supported in part by the Department of Energy, Contract DE-AC02-76SF00515.

\end{document}